\documentclass[aip,jcp,reprint,superscriptaddress,floatfix,longbibliography]{revtex4-1}
\usepackage{epsfig}
\usepackage{graphicx}
\usepackage{color}
\usepackage{dcolumn} 
\usepackage{bm} 
\usepackage{amssymb}
\usepackage[version=4]{mhchem}

\newcommand{\da}{\downarrow}
\newcommand{\ua}{\uparrow}

\begin{document}


\title{Fast simulation of soft x-ray near-edge spectra using a relativistic state-interaction approach: Application to closed-shell transition metal complexes}

\author{Sarah Pak}
\affiliation{Department of Chemistry, University of Memphis, Memphis, TN 38152, USA}
\author{Muhammed A. Dada}
\affiliation{Department of Chemistry, University of Memphis, Memphis, TN 38152, USA}
\author{Niranjan Govind}
\affiliation{Physical and Computational Sciences Directorate, 
         Pacific Northwest National Laboratory,
         Richland, WA 99352, USA}
\affiliation{Department of Chemistry, University of Washington, Seattle, WA 98195, USA}    
\author{Daniel R. Nascimento}
\email{daniel.nascimento@memphis.edu}
\affiliation{Department of Chemistry, University of Memphis, Memphis, TN 38152, USA}




\begin{abstract}

Spectroscopic techniques based on core-level excitations provide powerful tools for probing molecular and electronic structures with high spatial resolution. However, accurately calculating spectral features at the L or M edges is challenging due to the significant influence of spin-orbit and multiplet effects. While scalar-relativistic effects can be incorporated at minimal computational cost, accounting for spin-orbit interactions requires more complex computational frameworks. In this work, we develop and apply the state-interaction approach, incorporating relativistic effects using the ZORA-Kohn-Sham Hamiltonian, to simulate near-edge soft X-ray absorption spectra for closed-shell transition metal complexes. The computed spin-orbit splittings closely match those obtained from more rigorous methods. This approach provides a practical and cost-effective alternative to more rigorous two-component methods, making it particularly valuable for large-scale calculations and applications such as resonant inelastic X-ray scattering simulations, where capturing a large number of excited states is essential.
\end{abstract}

\maketitle

\section{Introduction}
\label{IntroSection}
Spectroscopic techniques based on core-level excitations offer powerful tools for probing molecular and electronic structures with high spatial resolution and atomic specificity. Over the past decades, advancements in free-electron laser (FEL) technologies \cite{Ackermann:2007:336,Emma:2010:641,huang2021features} have made these techniques more widely accessible, enabling researchers to tackle a broad range of scientific questions \cite{doi:10.1126/sciadv.adp0841,li2024attosecond,lin2024structural,driver2024attosecond,guo2024experimental,hutchison2023optical,bhowmick2023structural,leshchev2023revealing,larsen2024metal}.

Experiments at FEL facilities produce vast amounts of data that require thorough analysis and interpretation using theoretical models and computational simulations. However, advanced quantum chemistry methods, such as coupled-cluster theory, are computationally prohibitive for complex systems in realistic environments. 
Therefore, developing practical approaches that preserve essential physics while reducing computational costs—without compromising predictive accuracy—is crucial for advancing the field.

Density functional theory (DFT) based response approaches have proven highly effective in this context, offering a balance between accuracy and computational efficiency. For example, linear-response (LR) time-dependent density functional theory (TDDFT) \cite{casida1995recent,casida2009time} and the Tamm-Dancoff Approximation (TDA) have been widely employed in the computation of X-ray spectra of molecules and solids \cite{li2020real,bokarev2020theoretical,nascimento2022computational}, achieving remarkable success.

Accurate computation of spectral features at the L, M, or lower-energy edges, however, is significantly more complex due to the strong influence of spin-orbit (SO) coupling. In such cases, standard non-relativistic quantum chemistry approaches alone are insufficient to capture the relevant physics. Instead, computations based on the real-time propagation of the Dirac--Kohn--Sham density matrix (RT-DKS) \cite{kadek2015x,ye2022self}, 4-component (4c) Damped Response (DR) TDDFT \cite{konecny2021accurate}, LR\cite{stetina2019modeling} and real-time (RT)\cite{kasper2018modeling} exact two-component relativistic (X2C) TDDFT, and relativistic two-component zeroth-order regular approximation (ZORA) TDDFT \cite{stener2003time,wang2005calculation,fronzoni2005spin,casarin2007spin,fronzoni2012l2,hua2013fe} offer practical alternatives.

While scalar-relativistic (SR) effects can be seamlessly integrated into non-relativistic quantum chemistry codes with minimal impact on computational performance \cite{stoll2002relativistic,hirao2004recent,dolg2012relativistic,mosyagin2016generalized,liu2021relativistic}, SO interactions break spin symmetries, necessitating complex-valued spin-generalized frameworks that accommodate spin non-collinearity. The resulting algorithms can be up to 8$\times$ more expensive than in the real-valued, spin-restricted case for ground-state DFT, and up to 32$\times$ more expensive for LR-TDDFT.

In core-level spectroscopies, SO effects are significant even for relatively light elements. For example, the L$_{2,3}$ edge splittings of first-row transition metals (TMs) range from 5 to 10 eV --- comparable to core-hole lifetime broadening --- which causes overlap between the L$_2$ and L$_3$ features, making peak assignments very challenging. A similar trend is observed at the M and N edges of second and third-row TMs, respectively. While these shallow edges, accessible via soft X-rays, have historically been of little interest, current resonant-inelastic X-ray scattering (RIXS) experiments routinely probe them to gain deeper insights into the coupling between the core and valence excited states and the bonding characteristics of TM complexes \cite{jay2022capturing,lee2023ab,hahn2018probing,nascimento2021resonant,van2024correlating,Biasin:2021:Revealing,poulter2023uncovering,larsen2024metal}.

Here, we employ a linear-response treatment of a ZORA-Kohn-Sham (ZKS) Hamiltonian based on the relativistic model potential of van W{\"u}llen \cite{van1998molecular,van2005accurate}, to simulate the soft 
X-ray edge splittings of bare closed-shell transition metal cations and the near-edge X-ray absorption spectra of a series of closed-shell cyanometallates using the state-interaction ZKS approach with results that are comparable to state-of-the-art 4-component DKS and X2C methods.
We also demonstrate that, in the present scenario, spin-orbit coupling effects can be efficiently introduced on top of a scalar-relativistic reference via the state-interaction approach. The spectra of a series of cyanometallates obtained with the present approach is shown to closely reproduce those generated by the fully variational X2C-LR-TDDFT method, at a significantly lower computational cost.

The paper is organized as follows: In Section \ref{TheorySection} we describe the theory and working equations, the implementation is described in Section \ref{ImplSection}, the computational details are given in Section \ref{CompSection}, the results and discussion in Section \ref{ResultsSection}, and finally a conclusion in Section \ref{ConclusionSection}.

\section{Theory}
\label{TheorySection}
\subsection{Background}
The ground-state ZORA-Kohn-Sham (ZKS) problem is defined as
\begin{equation}
    \label{EQN:ZKS_eq}
    \hat{h}^{\rm ZKS} \varphi_i = \epsilon_i \varphi_i
\end{equation}
with Hamiltonian given (in atomic units) by
\begin{equation}
    \label{EQN:ZKS}
    \hat{h}^{\rm ZKS} = {\frac{\bm{p}^{2}}{2}}+{\bm{p}\left(\frac{\kappa-1}{2}\right)\bm{p}} + {\frac{\kappa^2}{4c^{2}}\bm{\sigma}\cdot (\nabla v^{\rm KS}\times \bm{p})} + v^{\rm KS}.
\end{equation}
Here, $\bm{p}$ is the momentum operator, $\bm{\sigma}$ is a vector of Pauli spin matrices, and $\kappa$ has the form
\begin{equation}
    \label{EQN:kappa}
    \kappa = \left [ 1 - \frac{v^{\rm KS}}{2c^2} \right ]^{-1}.
\end{equation}
$v^{\rm KS}$ is the usual Kohn--Sham (KS) potential accounting for the external nuclear, Coulomb repulsion, and exchange-correlation potentials:
\begin{equation}
    v^{\rm KS}(\bm{r}) = - \sum_{A} \frac{Z_A}{|\bm{r} - \bm{R_A}|} + \int \frac{\rho(\bm{r}')}{|\bm{r} - \bm{r}'|} d\bm{r}' + \frac{\delta E_{\rm xc}[\rho]}{\delta \rho(\bm{r})}.
\end{equation}

The terms in the right-hand side of Eq. \ref{EQN:ZKS} can be easily identified as the classical kinetic energy, a scalar relativistic correction to the classical kinetic energy, the spin-orbit interaction, and the KS potential terms, respectively. 

Direct application of the ZKS Hamiltonian in the form of Eq. \ref{EQN:ZKS}, however, is known to be problematic as the dependence of the relativistic terms on the KS potential can lead to convergence and gauge-invariance issues \cite{van1998molecular}.

To bypass this problem and facilitate the evaluation of analytical geometry derivatives, van W{\"u}llen \cite{van1998molecular} proposed to replace the KS potential appearing in the relativistic terms (but not the last term in eq. \ref{EQN:ZKS}) by an atom-based effective potential of the form
\begin{align}
    \label{EQN:potential}
    v^{\rm eff}(\bm{r}) = &- \sum_{A} \frac{Z_A}{|\bm{r} - \bm{R_A}|} + \frac{\delta E_{LDA}[\tilde{\rho}]}{\delta \tilde{\rho}(\bm r)} \\ \nonumber &+ \frac{2}{\sqrt{\pi}} \sum_{A,i} c_{i}^{A} \sqrt{\alpha_{i}^{A}}\times F_0 \left ( \alpha_{i}^{A} (\bm{r} - \bm{R_A})^2 \right ),
\end{align}
where, $F_0$ denotes the Boys function, and $c_{i}^{A}$ and $\alpha_{i}^{A}$ are the coefficients of the $i$th $s$-type Gaussian function centered on atom $A$, defining a model potential (provided in the SI). This model potential has the same components as the KS potential, however, the Coulomb and exchange-correlation terms are now evaluated with respect to a model density $\tilde{\rho}$ expressed as
\begin{equation}
    \tilde{\rho}(\bm{r}) = \pi^{-3/2} \sum_{iA} c_i^A (\alpha_i^A)^{3/2} \exp(-\alpha_i^A |\bm{r} - \bm{R}_A|^2).
\end{equation}
Note that in this effective potential, the exchange-correlation functional has been fixed as the local-density approximation (LDA) regardless of the exchange-correlation functional chosen for the SCF procedure. The obvious advantages of utilizing an effective potential of this form are the fact that it needs to be evaluated only once before the SCF cycle, and the ease to evaluate analytical derivatives, thus reducing the computational complexity of the overall procedure. In the present work, we adopt a slight modification by excluding the exchange-correlation term within the ZORA atomic model potential, while retaining it in the last term of Eq. \ref{EQN:ZKS}. Previous work\cite{van2000gradients,faas2000ab,van2006note,Nichols:2009:491,
nascimento2022computational} has shown that this term has negligible effect on ground and excited states. 

Once the ZKS eigenpairs have been obtained self-consistently, we employ the scaled ZORA procedure of van Lenthe {\em et al.}  to correct the {\em occupied} spinor energies up to first-order in the regular expansion \cite{van-Lenthe:1994:9783}:
\begin{equation}
    \tilde{\epsilon}_i = \left(1 + \langle \varphi_i| \bm{\sigma}\cdot\bm{p} \frac{c^2}{(2c^2 - v^{\rm eff})^2} \bm{\sigma}\cdot\bm{p} |\varphi_i \rangle \right)^{-1} \epsilon_i. 
\end{equation}
These energies are a better approximation to the 4-component DKS energies, thus providing a better starting point for our linear-response computations.

Finally, the linear-response TDDFT/TDA problem is solved employing the core-valence separation approach to obtain quasi-relativistic core-level excitation energies and oscillator strengths. Within the TDA,

\begin{equation}
  \label{EQN:TDDFT}
    AX = \Omega X
\end{equation}
with 
\begin{equation}
    A_{iajb} = (\epsilon_a  - \tilde{\epsilon}_i)\delta_{ij} \delta_{ab} + g_{iabj} - \alpha g_{ijba} + f^{\rm xc}_{iabj}(\alpha).
\end{equation}
Here, $f^{\rm xc}(\alpha)$ denotes the scaled exchange-correlation kernel, where $\alpha$ is the hybrid parameter, and $g_{pqrs}$ are electron repulsion integrals defined, respectively, as
\begin{equation}
    f^{\rm xc}_{pqrs}(\alpha) =  \langle \varphi_p \varphi_r | \left [ (1-\alpha) \frac{\delta^2 E_{\rm x}}{\delta\rho^2} + \frac{\delta^2 E_{\rm c}}{\delta\rho^2} \right ]| \varphi_q \varphi_s \rangle 
\end{equation}
and
\begin{equation}
    g_{pqrs} = \langle \varphi_p \varphi_r |\frac{1}{|\bm{r} - \bm{r}'|}| \varphi_q \varphi_s \rangle. 
\end{equation}

Since the ZKS hamiltonian now includes spin-orbit interactions, spin is no longer a good quantum number and the TDDFT/TDA matrix will now contain spin-flip blocks with complex-valued entries. That is,
\begin{equation}
    A = 
    \begin{pmatrix}
      A_{\ua \ua \ua \ua} & A_{\ua \ua \ua \da} &  A_{\ua \ua \da \ua} & A_{\ua \ua \da \da} \\
      A_{\ua \da \ua \ua} & A_{\ua \da \ua \da} &  A_{\ua \da \da \ua} & A_{\ua \da \da \da} \\
      A_{\da \ua \ua \ua} & A_{\da \ua \ua \da} &  A_{\da \ua \da \ua} & A_{\da \ua \da \da} \\
      A_{\da \da \ua \ua} & A_{\da \da \ua \da} &  A_{\da \da \da \ua} & A_{\da \da \da \da}
    \end{pmatrix}, ~ A \in \mathbb{C}.
\end{equation}

This matrix has 8$\times$ as many elements as in the real-valued unrestricted case, where only the  $A_{\ua \ua \ua \ua}$, $A_{\ua \ua \da \da}$, $A_{\da \da \ua \ua}$, and $A_{\da \da \da \da}$ blocks are non-zero, and 32$\times$ as many elements as in the real-valued restricted case, where only one block is needed. The construction of the spin-flip blocks strictly requires a non-collinear exchange-correlation functional, which we do not consider in this work. 

Here, we consider two approximations where $A$ is constructed with and without $f^{\rm xc}(\alpha)$, while using a standard collinear functional for the ground state orbitals. The latter stems from a recent study by some of the authors, which 
demonstrated that $f^{\rm xc}(\alpha)$ has no significant effect in the calculation of core-level spectra in transition metal complexes.~\cite{pak2024role} This approximation resembles a {\em scaled} configuration interaction singles (CIS) approach performed with approximate ZKS spinors. We refer to these as ZKS/TDA and ZKS/SCIS, respectively. SO interactions are introduced by using a state-interaction approach described below.~\cite{malmqvist1989casscf}




\subsection{State-Interaction Approach}
The state-interaction approach allows us to couple a set of states $\{|I\rangle\}$ via an operator $\hat{V}$ by direct diagonalization of the matrix representation of $\hat{V}$ in the basis spanned by $\{|I\rangle\}$. In the present context, this set of states is a subset of the solutions to the full scalar-relativistic (SR) TDA equations obtained by removing the SO term in the ZKS Hamiltonian, and $\hat{V}$ is the SO-coupling operator. That is, we obtain SO-coupled excitation energies $\Xi$ and states $Z$, by solving 
\begin{equation}
    \mathcal{H}^{\rm SO} Z = \Xi {Z},
\end{equation}
with
\begin{equation}
    \mathcal{H}^{\rm SO}_{IJ} = \Omega_I \delta_{IJ} + \langle I|{\frac{\kappa^2}{4c^{2}}\bm{\sigma}\cdot (\nabla v^{\rm eff}\times \bm{p})} |J\rangle,
\end{equation}
where $\Omega_I$ and $|I\rangle$ are related to the solutions of the scalar-relativistic TDA problem as 
\begin{equation}
  \label{EQN:TDA_SR}
    A^{\rm SR} X = \Omega X, \text{ and } |I\rangle = \sum_{ia} X^I_{ia} \hat{a}^\dagger_a \hat{a}_i |\Phi_0\rangle.
\end{equation}
Here, $|\Phi_0\rangle$ represents the reference KS determinant and indices $i$ and $a$ span occupied and virtual orbitals, respectively.

An immediate advantage of the state-interaction approach is that the scalar relativistic states can be obtained taking full advantage of real and spin symmetries. Furthermore, since $\mathcal{H}^{\rm SO}$ is constructed from a subset of $\{|I\rangle\}$, its dimension is only a fraction of the dimension of $A^{\rm SR}$, making Eq. \ref{EQN:TDA_SR} the most expensive part of the procedure. Altogether, the approach outlined above leads to reduction in the computational cost by factors of 8$\times$ and 32$\times$ for unrestricted and restricted SR references, respectively. We will refer to these approaches as SR-ZKS/SCIS+SO, when $A^{\rm SR}$ is constructed without $f^{\rm xc}(\alpha)$, and SR-ZKS/TDA+SO when $A^{\rm SR}$ is constructed with a standard collinear $f^{\rm xc}(\alpha)$.

\section{Implementation}
\label{ImplSection}
Solving Eq. \ref{EQN:ZKS_eq} requires evaluation of the ZORA integrals. These integrals are not standard in electronic structure theory, but can be easily evaluated in an atomic orbital basis set, $\{\chi_\mu\}$, with knowledge of the basis functions gradient, $\nabla \chi_\mu$, and the effective potential, $v^{\rm eff}$ evaluated using a numerical grid.

The scalar-relativistic kinetic energy integral takes the form
\begin{align}
     T^{\rm SR}_{\mu \nu} &= \int  \chi_\mu^\dagger ( \bm{r}) \left [ {\frac{\bm{p}^{2}}{2}} + {\bm{p}\left(\frac{\kappa-1}{2}\right)\bm{p}} \right ] \chi_\nu ( \bm{r}) d \bm{r} \\
     &= \int \frac{c^2}{2c^2 - v^{\rm eff}(\bm{r})}  \nabla \chi_\mu^\dagger ( \bm{r}) \cdot \nabla \chi_\nu( \bm{r}) d \bm{r} \label{EQN:TSR_MATRIX}
\end{align}
while the SO integral is given by
\begin{align}
     H^{\rm SO}_{\mu \nu} &= \int  \chi_\mu^\dagger ( \bm{r}) \left [ {\frac{\kappa^2}{4c^{2}}\bm{\sigma}\cdot \nabla v^{\rm eff}\times \bm{p}} \right ] \chi_\nu ( \bm{r}) d \bm{r} \\
     &= {\bm{\sigma}} \cdot \int \frac{v^{\rm eff} (\bm{r})}{4c^2 - 2v^{\rm eff}(\bm{r})} 
     \nabla \chi_\mu^\dagger ( \bm{r}) \times \nabla \chi_\nu( \bm{r}) d \bm{r} \label{EQN:SO_MATRIX}.
\end{align}
Note that since the operator in Eq. \ref{EQN:TSR_MATRIX} does not have a spin component, $T^{\rm SR}_{\mu \nu}$ will have a block-diagonal structure, facilitating its incorporation into standard non-relativistic electronic structure codes.

For convenience, it is useful to separate the spin and spatial components of $H^{\rm SO}_{\mu \nu}$ and label the components of the integral in the right-hand side of Eq. \ref{EQN:SO_MATRIX} explicitly as
\begin{align}
     h^{x}_{\mu \nu} = \int \frac{v^{\rm eff} (\bm{r})}{4c^2 - 2v^{\rm eff} (\bm{r})} 
     \left [ \frac{\partial \chi_\mu^\dagger }{\partial y} \frac{\partial \chi_\nu }{\partial z} -  \frac{\partial \chi_\mu^\dagger }{\partial z} \frac{\partial \chi_\nu }{\partial y}  \right ] d \bm{r} \\
     h^{y}_{\mu \nu} = \int \frac{v^{\rm eff} (\bm{r})}{4c^2 - 2v^{\rm eff} (\bm{r})} 
     \left [ \frac{\partial \chi_\mu^\dagger }{\partial z} \frac{\partial \chi_\nu }{\partial x} -  \frac{\partial \chi_\mu^\dagger }{\partial x} \frac{\partial \chi_\nu }{\partial z}  \right ] d \bm{r}
     \\
     h^{z}_{\mu \nu} = \int \frac{v^{\rm eff} (\bm{r})}{4c^2 - 2v^{\rm eff} (\bm{r})} 
     \left [ \frac{\partial \chi_\mu^\dagger }{\partial x} \frac{\partial \chi_\nu }{\partial y} -  \frac{\partial \chi_\mu^\dagger }{\partial y} \frac{\partial \chi_\nu }{\partial x}  \right ] d \bm{r},
\end{align}
where the spatial dependence of the atomic orbitals is implied. Upon taking the dot product with ${\bm{\sigma}}$, $H^{\rm SO}$ assumes the final form
\begin{equation}
    H^{\rm SO} = 
    \begin{pmatrix}
      h^z & h^x - ih^y \\
      h^x + ih^y & -h^z
    \end{pmatrix}.
\end{equation}

$H^{\rm SO}$ is complex-valued, and will lead to spin-mixing, imposing the need for a complex-valued generalized self-consistent-field procedure to solve Eq. \ref{EQN:ZKS_eq}. However, as discussed in Section II B, this need can be bypassed by employing the state-interaction approach, which provides significant savings, especially when dealing with closed-shell systems. For these systems, the $A^{\rm SR}$ matrix can be spin-adapted to yield singlet and restricted triplet components:
\begin{align}
    \label{EQN:SR_TDA_0}
    A_{\rm SR}^{\rm 0,0} S &= \Omega^{\rm 0,0} S \\
    \label{EQN:SR_TDA_1}
    A_{\rm SR}^{\rm 1,0} T &= \Omega^{\rm 1,0} T ,
\end{align}
where
\begin{align}
        A_{iajb}^{0,0} &= (\epsilon_a  - \tilde{\epsilon}_i)\delta_{ij} \delta_{ab} + 2g_{iabj} -\alpha g_{ijba} + f^{\rm xc}_{iabj}(\alpha) \\
        A_{iajb}^{1,0} &= (\epsilon_a  - \tilde{\epsilon}_i)\delta_{ij} \delta_{ab} -\alpha g_{ijba} + f^{\rm xc}_{iabj}(\alpha).
\end{align}
The real-valued solution vectors are then used to construct a 2-component complex spin basis, $\{|S,M_S\rangle \}$, in which the spin-orbit Hamiltonian will be evaluated:
\begin{equation}
     |0,0 \rangle = \frac{1}{\sqrt{2}}
     \begin{pmatrix}
      S \\
      S
    \end{pmatrix}
    ~ \text{and} ~ 
    |1,0 \rangle = \frac{1}{\sqrt{2}}
     \begin{pmatrix}
       T \\
      -T
    \end{pmatrix}.
\end{equation}
The remaining triplet basis are obtained, without additional computation, through the action of raising and lowering spin operators on $|1,0 \rangle$:
\begin{equation}
     |1,+1 \rangle = \frac{1}{\sqrt{2}} \left (\sigma_x + i \sigma_y \right )|1,0 \rangle = 
     \begin{pmatrix}
     0 \\
     T
    \end{pmatrix}
\end{equation}
and
\begin{equation}
     |1,-1 \rangle = \frac{1}{\sqrt{2}} \left (\sigma_x - i \sigma_y \right )|1,0 \rangle = 
     \begin{pmatrix}
     -T \\
     0
    \end{pmatrix}
\end{equation}

The block-structure of the spin-orbit Hamiltonian in the basis spanned by $ |S,M_S\rangle =  |0,0\rangle \otimes |1,-1\rangle \otimes |1,0\rangle \otimes |1,+1\rangle $ is then given by
\begin{equation}
     \mathcal{H}_{\rm SO} =  
     \begin{pmatrix}
     \Omega^{0,0} & \frac{\sqrt{2}}{2} h^{+1}_{\rm ST} & h^0_{\rm ST} & -\frac{\sqrt{2}}{2} h^{-1}_{\rm ST} \\
     \frac{\sqrt{2}}{2} h^{-1}_{\rm TS} & \Omega^{1,0} + h^0_{\rm TT} &  -\frac{\sqrt{2}}{2} h^{-1}_{\rm TT}  & 0\\
     h^0_{\rm TS} & -\frac{\sqrt{2}}{2} h^{+1}_{\rm TT} & \Omega^{1,0} & -\frac{\sqrt{2}}{2} h^{-1}_{\rm TT} \\
     -\frac{\sqrt{2}}{2} h^{+1}_{\rm TS} & 0 & -\frac{\sqrt{2}}{2} h^{+1}_{\rm TT} & \Omega^{1,0}-h^0_{\rm TT}
    \end{pmatrix}
\end{equation}
with
\begin{align}
    [h_{\rm ST}^{\xi}]_{IJ} = \sum_{ia} S_{ia}^{I} \sum_{jb} \left (h_{ab}^{\xi} \delta_{ij} - h_{ji}^{\xi}
    \delta_{ab} \right) T_{jb}^{J} \\
    [h_{\rm TS}^{\xi}]_{IJ} = \sum_{ia} T_{ia}^{I} \sum_{jb} \left (h_{ab}^{\xi} \delta_{ij} - h_{ji}^{\xi}
    \delta_{ab} \right) S_{jb}^{J} \\
    [h_{\rm TT}^{\xi}]_{IJ} = \sum_{ia} T_{ia}^{I} \sum_{j,b} \left (h_{ab}^{\xi} \delta_{ij} + h_{ji}^{\xi}
    \delta_{ab} \right) T_{jb}^{J}.
\end{align}
Here, $\xi \in \{-1,0,1 \}$ denote the components of the spin-orbit coupling matrix, given explicitly, in the canonical basis as
\begin{align}
    h^0_{pq} &= \sum_{\mu \nu} C_{\mu p} h^z_{\mu \nu} C_{\nu q} \\
    h^{+1}_{pq} &= \sum_{\mu \nu} C_{\mu p} \left ( h^x_{\mu \nu} + ih^y_{\mu \nu} \right ) C_{\nu q} \\
    h^{-1}_{pq} &= \sum_{\mu \nu} C_{\mu p} \left ( h^x_{\mu \nu} - ih^y_{\mu \nu} \right ) C_{\nu q}.
\end{align}

Finally, $\mathcal{H}_{\rm SO}$ can be diagonalized and the solution vectors, $Z$, can be used to evaluate transition dipole moments as
\begin{equation}
    \bm{\mu}^n = \sqrt{2} \sum_{ia}  \bm{\mu}_{ia}\sum_I S_{ia}^{I} Z_{I}^n,
\end{equation}
where $I$ and $n$ denote uncoupled and coupled states, respectively.

\section{Computational Details}
\label{CompSection}
Geometries were optimized in the gas phase using the NWChem software package \cite{Valiev:2010:1477,Apra:2012:184102}, and employed the PBE0 \cite{Perdew:1996:9982,Adamo:1999:6158} exchange-correlation functional paired with the 6-31G* \cite{ditchfield1971a,hehre1972a,hariharan1973a} basis set for C and N, and the Sapporo-DKH3-DZP-2020-diffuse \cite{Noro:2012:1124} basis set for the metal center. Scalar relativistic effects were taken into account by means of the SR-ZORA correction as implemented in NWChem \cite{Nichols:2009:491}, and point-group symmetry was enforced to speed-up optimizations. The resulting optimal geometric parameters are shown in Table \ref{TAB:cyanometallates}. 
\begin{table}[htpb!]
    \centering
    {\renewcommand{\arraystretch}{1.5}
    \begin{tabular}{|c|c|c|c|}
    \hline
    \hline
    Point Group & Complex & $r(M-C)$ & $r(M-N)$  \\
    \hline
    $O_h$ & \ce{[Cr(CN)6]^{6-}} & 2.08 & 3.29 \\
          & \ce{[Mo(CN)6]^{6-}} & 2.13 & 3.24 \\
          & \ce{[W(CN)6]^{6-}}  & 2.19 & 3.41 \\
          & \ce{[Fe(CN)6]^{4-}} & 1.97 & 3.15 \\
          & \ce{[Ru(CN)6]^{4-}} & 2.06 & 3.25 \\
          & \ce{[Os(CN)6]^{4-}} & 2.03 & 3.22 \\
          \hline
    $D_{2h}$ & \ce{[Cu(CN)2]^{-}} & 1.80 & 2.97 \\
          & \ce{[Ag(CN)2]^{-}} & 1.83 & 3.28 \\
          & \ce{[Au(CN)2]^{-}} & 1.80 & 3.24 \\
    \hline
    \hline
    \end{tabular}}
    \caption{Optimized geometrical parameters for the cyanometallates relevant to the present work. Atomic separations are shown in \AA.}
    \label{TAB:cyanometallates}
\end{table}

Excited-state calculations were performed employing the PBE0 exchange-correlation functional and the Dyall-v2z \cite{dyall2007a,dyall2023b} primitive basis set. The SR-ZKS/SCIS+SO, SR-ZKS/TDA+SO, and the fully 2-component SO-ZKS/SCIS methods were implemented as an {\em in-house} python code interfaced with the PySCF \cite{sun2018pyscf} quantum chemistry package, while reference DKS and X2C calculations were performed using the ReSpect code \cite{repisky2020respect}.

\section{Results and Discussion}
\label{ResultsSection}
\subsection{Bare Transition Metal Cations}

\begin{figure}[htb!]
    \includegraphics[scale=1]{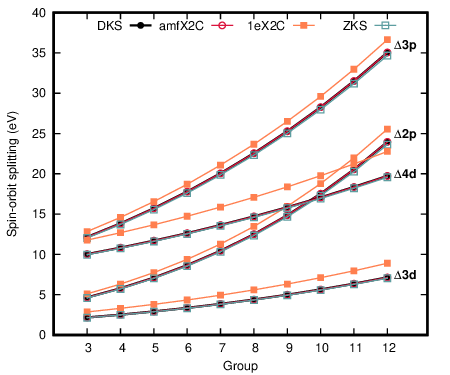}
    \caption{ SO splittings in the Kohn--Sham orbital energies for a series of bare, d$^0$ transition metal cations, computed with different relativistic Hamiltonians. Here, $\Delta$2p, $\Delta$3p/3d, and $\Delta$4d, are computed for atoms in the 4th, 5th, and 6th periods of the periodic table, respectively.}
    \label{FIG:mo_splittings}
\end{figure}

In order to understand how well the modified ZORA effective potential described earlier is able to reproduce the Dirac--Kohn--Sham (DKS) Hamiltonian, we begin by calculating the SO splittings in the ZKS eigenvalues of a series of bare transition metal cations (Figure \ref{FIG:mo_splittings}). Here, we focus on core-level orbitals with SO splittings between 2 and 40 eV.
These are generally orbitals with principal quantum number $(N-2)$, with $N$ corresponding to the period in which the atom appears in the periodic table.    

As shown in Figure \ref{FIG:mo_splittings}, the SO splittings computed with the atomic-mean-field exact 2-component method (amfX2C) of Konecny {\em et al.}\cite{konecny2023exact} and the effective potential ZKS Hamiltonians, are essentially identical to the ones obtained with the more computationally expensive DKS Hamiltonian. The only Hamiltonian to perform poorly was the 1-electron X2C (1eX2C) Hamiltonian \cite{konecny2023exact}, where relativistic 2-electron contributions are completely neglected. These results highlight the importance of accounting for two-electron relativistic corrections, even if it is done by means of an effective potential. 

\begin{figure}[htb!]
    \includegraphics[scale=1]{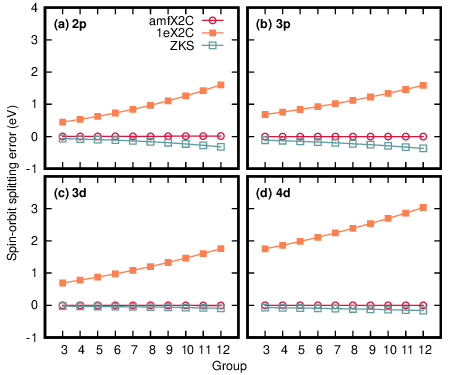}
    \caption{Error in the SO splittings in the Kohn--Sham orbital energies for a series of bare, d$^0$ transition metal cations, computed with different relativistic Hamiltonians.}
    \label{FIG:mo_splittings_error}
\end{figure}

To better quantify the error in the SO splittings, we show the difference between the SO splitting obtained with the amfX2C, 1eX2C, and ZKS methods against those obtained with the reference DKS method in Figure \ref{FIG:mo_splittings_error}.
As one can observe, the error tends to increase monotonically across a period, but decreases across a group, with the largest absolute error for the ZKS method being 0.37 eV ($\approx$ 1\%) for the \ce{Cd^{12+}} 3p orbitals. The average (signed) error for ZKS across a period was evaluated to be -0.17, -0.22, -0.06, and -0.11 eV, for the 2p, 3p, 3d, and 4d orbitals, respectively. In contrast, the errors for 1eX2C method increase both across periods and groups, with the largest error reaching 3.04 eV ($>$15\%) for the \ce{Hg^{12+}} 4d orbitals. For 1eX2C, the average errors were 0.95 (2p), 1.10 (3p), 1.18 (3d), and 2.35 eV (4d). The SO splittings for each cation calculated with different Hamiltonians are presented in Table \ref{TAB:mo_splittings}. 

\begin{table*}[htb!]
\centering
\setlength{\tabcolsep}{0.80em} 
{\renewcommand{\arraystretch}{1.5}
\begin{tabular}{|c |c c c c c c c c c c c c |}
\hline
\hline
$\Delta$2p & \ce{Sc^3+} & \ce{Ti^4+} & \ce{V^5+} & \ce{Cr^6+} & \ce{Mn^7+} & \ce{Fe^8+} & \ce{Co^9+} & \ce{Ni^10+} & \ce{Cu^11+} & \ce{Zn^12+} & MSD & RMSD \\
\hline
DKS &  4.66 &  5.79 &  7.11 &  8.66 & 10.44 & 12.50 & 14.85 & 17.52 & 20.54 & 23.95 & & \\
amfX2C &  4.66 &  5.79 &  7.11 &  8.66 & 10.45 & 12.51 & 14.86 & 17.53 & 20.55 & 23.96 & \bf{ 0.01} & \bf{ 0.00} \\
1eX2C &  5.11 &  6.32 &  7.74 &  9.38 & 11.29 & 13.47 & 15.96 & 18.78 & 21.97 & 25.55 & \bf{ 0.95} & \bf{ 0.32} \\
ZKS &  4.59 &  5.71 &  7.02 &  8.54 & 10.31 & 12.34 & 14.65 & 17.29 & 20.27 & 23.62 & \bf{-0.17} & \bf{ 0.06} \\
\hline
\hline
$\Delta$3p & \ce{Y^3+} & \ce{Zr^4+} & \ce{Nb^5+} & \ce{Mo^6+} & \ce{Tc^7+} & \ce{Ru^8+} & \ce{Rh^9+} & \ce{Pd^10+} & \ce{Ag^11+} & \ce{Cd^12+} & MSD & RMSD \\
\hline

DKS & 12.14 & 13.84 & 15.71 & 17.78 & 20.05 & 22.55 & 25.28 & 28.26 & 31.51 & 35.05 & & \\
amfX2C & 12.14 & 13.84 & 15.71 & 17.78 & 20.05 & 22.55 & 25.28 & 28.26 & 31.51 & 35.04 & \bf{-0.00} & \bf{ 0.00} \\
1eX2C & 12.83 & 14.60 & 16.55 & 18.71 & 21.07 & 23.67 & 26.51 & 29.60 & 32.97 & 36.63 & \bf{ 1.10} & \bf{ 0.36} \\
ZKS & 12.03 & 13.71 & 15.56 & 17.61 & 19.86 & 22.33 & 25.03 & 27.97 & 31.18 & 34.68 & \bf{-0.22} & \bf{ 0.07} \\
\hline
\hline
$\Delta$3d & \ce{Y^3+} & \ce{Zr^4+} & \ce{Nb^5+} & \ce{Mo^6+} & \ce{Tc^7+} & \ce{Ru^8+} & \ce{Rh^9+} & \ce{Pd^10+} & \ce{Ag^11+} & \ce{Cd^12+} & MSD & RMSD \\
\hline

DKS &  2.17 &  2.52 &  2.92 &  3.36 &  3.85 &  4.39 &  4.99 &  5.64 &  6.36 &  7.14 & & \\
amfX2C &  2.17 &  2.52 &  2.92 &  3.36 &  3.85 &  4.39 &  4.99 &  5.64 &  6.35 &  7.13 & \bf{-0.00} & \bf{ 0.00} \\
1eX2C &  2.86 &  3.30 &  3.79 &  4.34 &  4.94 &  5.60 &  6.32 &  7.11 &  7.97 &  8.90 & \bf{ 1.18} & \bf{ 1.18} \\
ZKS &  2.14 &  2.49 &  2.88 &  3.32 &  3.80 &  4.34 &  4.93 &  5.57 &  6.28 &  7.05 & \bf{-0.06} & \bf{ 0.02} \\
\hline
\hline
$\Delta$4d & \ce{Lu^3+} & \ce{Hf^4+} & \ce{Ta^5+} & \ce{W^6+} & \ce{Re^7+} & \ce{Os^8+} & \ce{Ir^9+} & \ce{Pt^10+} & \ce{Au^11+} & \ce{Hg^12+} & MSD & RMSD \\
\hline

DKS & 10.02 & 10.83 & 11.69 & 12.63 & 13.63 & 14.70 & 15.85 & 17.07 & 18.36 & 19.73 & & \\
amfX2C & 10.02 & 10.83 & 11.69 & 12.63 & 13.63 & 14.70 & 15.85 & 17.06 & 18.36 & 19.73 & \bf{-0.00} & \bf{ 0.00} \\
1eX2C & 11.77 & 12.69 & 13.68 & 14.74 & 15.88 & 17.09 & 18.38 & 19.76 & 21.22 & 22.77 & \bf{ 2.35} & \bf{ 0.75} \\
ZKS &  9.95 & 10.75 & 11.61 & 12.54 & 13.53 & 14.59 & 15.72 & 16.93 & 18.21 & 19.57 & \bf{-0.11} & \bf{ 0.04} \\
\hline
\hline
\end{tabular}}
\caption{SO splittings, mean signed deviation (MSD), and root mean squared deviation (RMSD) in the Kohn--Sham orbital energies for a series of bare, d$^0$ transition metal cations, computed with different relativistic Hamiltonians.}
\label{TAB:mo_splittings}
\end{table*}

\begin{figure}[htb!]
    \includegraphics[scale=1]{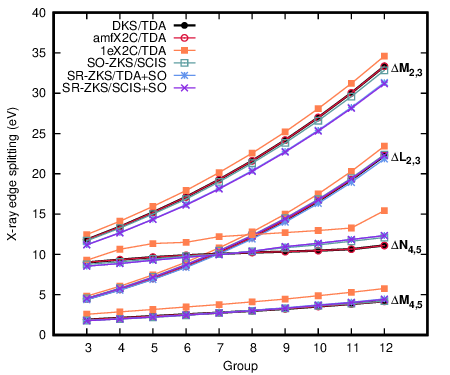}
    \caption{X-ray near-edge splittings due to spin-orbit coupling for a series of bare, d$^0$ transition metal cations, computed with different relativistic Hamiltonians and approximations. The splittings are calculated with respect to the spectral features with dominant p$_{3/2} \leftrightarrow$ d$_{5/2}$ and p$_{1/2} \leftrightarrow$ d$_{3/2}$ character.}
    \label{FIG:edge_splittings}
\end{figure}

Next, we analyze how the choice of relativistic approximation affects the spectral near-edge splittings. In Figure \ref{FIG:edge_splittings}, we report the splittings between excited states with dominant 2p$_{3/2}\to{\rm 3d}_{5/2}$ and 2p$_{1/2}\to{\rm 3d}_{3/2}$ character ($\Delta$L$_{2,3}$) for 4th-period cations, between excited states with dominant 3p$_{3/2}\to{\rm 4d}_{5/2}$ and 3p$_{1/2}\to{\rm 4d}_{3/2}$ ($\Delta$M$_{2,3}$), and 3d$_{3/2}\to{\rm 5p}_{1/2}$ and 3d$_{1/2}\to{\rm 5p}_{3/2}$ character ($\Delta$M$_{4,5}$) for 5th-period cations, and between excited states with dominant 4d$_{3/2}\to{\rm 6p}_{1/2}$ and 4d$_{5/2}\to{\rm 6p}_{3/2}$ character ($\Delta$N$_{4,5}$) for 6th-period cations. These excitations correspond to the dominant features in the soft X-ray absorption spectra.
The excited-state energies are calculated using both the spinor-based approach, where the ground-state ZKS equations are solved with variational inclusion of scalar relativistic {\em and} spin-orbit effects, followed by a linear-response calculation, and the state-interaction approach as outlined earlier. 

When comparing the spinor-based approaches, amfX2C/TDA reproduced the DKS/TDA results almost exactly, followed by the SO-ZKS/SCIS approach, which yields equally good results for the L$_{2,3}$, M$_{2,3}$, and M$_{4,5}$ splittings. For the N$_{4,5}$ splittings, the SO-ZKS/SCIS approach is indistinguishable from the reference (DKS/TDA) up to \ce{Fe^{8+}}, but deviations from the reference become pronounced beyond this point, reaching an error of almost 1 eV for \ce{Hg^{12+}}. One could think that this error is a manifestation of the scaled CIS approach for the excited states, but we demonstrate below that this is not the case. The performance of the 1eX2C/TDA method is overall poor, as expected given the large errors in the single-particle energies shown in Figure \ref{FIG:mo_splittings_error}, however, it is worth mentioning that approaches able to reduce this error have been reported in the literature \cite{ehrman2023improving}.

\begin{figure}[htbp!]
    \includegraphics[scale=1]{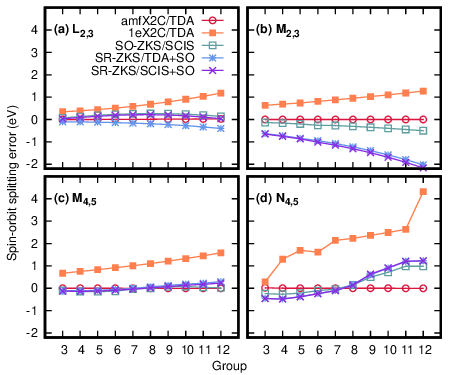}
    \caption{Error in the X-ray edge splittings for a series of bare, d$^0$ transition metal cations, computed with different relativistic Hamiltonians and approximations. The splittings are calculated with respect to the spectral features with dominant p$_{3/2} \leftrightarrow$ d$_{5/2}$ and p$_{1/2} \leftrightarrow$ d$_{3/2}$ character.}
    \label{FIG:edge_splittings_error}
\end{figure}

The two state-interaction approaches perform very similarly regardless of whether or not the exchange-correlation kernel is included in the construction of the TDA matrix. This observation is in corroboration with our previous study showing that the exchange-correlation kernel plays a negligible role in the calculation of core-level excited states \cite{pak2024role}, which also eliminates the removal of the exchange-correlation kernel in SO-ZKS/SCIS as a possible reason for the discrepancies in the N$_{4,5}$ splittings observed beyond \ce{Fe^{8+}}. Furthermore, the fact that SR-ZKS/TDA+SO and SO-ZKS/SCIS predict very close N$_{4,5}$ splittings, implies that the variational inclusion of SO effects during ground state optimization does not significantly affect these states either, leaving the parametrization of the ZORA effective potential as the likely source of error. In contrast, the M$_{2,3}$ splittings, are significantly affected by the inclusion of SO effects during the ground state optimization. Finally, the L$_{2,3}$ and M$_{4,5}$ predicted by all ZKS approaches are in very close agreement with those calculated using the DKS/TDA method. Upon closer inspection, one can observe that in fact, the L$_{2,3}$ are very slightly affected by the neglect of the exchange-correlation kernel in the TDA matrix causing the SR-ZKS/TDA+SO splittings to be slightly underestimated with respect to the DKS/TDA ones (-0.2 eV in average), while the SR-ZKS/SCIS+SO and SO-ZKS/SCIS splittings are slightly overestimated (0.2 eV in average). These differences are better visualized in Figure \ref{FIG:edge_splittings_error}, which reports the error in the edge splittings with respect to the DKS/TDA reference. The edge splittings for individual cations and corresponding error analysis are provided in Table \ref{TAB:edge_splittings}.

\begin{table*}[htbp!]
\centering
\setlength{\tabcolsep}{0.6em} 
{\renewcommand{\arraystretch}{1.5}
\begin{tabular}{|c |c c c c c c c c c c c c |}
\hline
\hline
$\Delta$L$_{2,3}$ & \ce{Sc^3+} & \ce{Ti^4+} & \ce{V^5+} & \ce{Cr^6+} & \ce{Mn^7+} & \ce{Fe^8+} & \ce{Co^9+} & \ce{Ni^10+} & \ce{Cu^11+} & \ce{Zn^12+} & MSD & RMSD \\
\hline
DKS/TDA &  4.48 &  5.66 &  7.01 &  8.52 & 10.21 & 12.10 & 14.22 & 16.60 & 19.26 & 22.25 & & \\
amfX2C/TDA &  4.48 &  5.67 &  7.01 &  8.53 & 10.22 & 12.12 & 14.24 & 16.62 & 19.29 & 22.28 & \bf{ 0.01} & \bf{ 0.00} \\
1eX2C/TDA &  4.82 &  6.05 &  7.45 &  9.03 & 10.80 & 12.79 & 15.01 & 17.50 & 20.30 & 23.43 & \bf{ 0.69} & \bf{ 0.23} \\
SO-ZKS/SCIS &  4.55 &  5.80 &  7.20 &  8.75 & 10.47 & 12.37 & 14.48 & 16.84 & 19.46 & 22.39 & \bf{ 0.20} & \bf{ 0.07} \\
SR-ZKS/TDA+SO &  4.38 &  5.56 &  6.89 &  8.38 & 10.05 & 11.91 & 14.00 & 16.32 & 18.93 & 21.85 & \bf{-0.21} & \bf{ 0.07} \\
SR-ZKS/SCIS+SO &  4.51 &  5.76 &  7.16 &  8.71 & 10.42 & 12.32 & 14.43 & 16.77 & 19.37 & 22.28 & \bf{ 0.14} & \bf{ 0.05} \\
\hline
\hline
$\Delta$M$_{2,3}$ & \ce{Y^3+} & \ce{Zr^4+} & \ce{Nb^5+} & \ce{Mo^6+} & \ce{Tc^7+} & \ce{Ru^8+} & \ce{Rh^9+} & \ce{Pd^10+} & \ce{Ag^11+} & \ce{Cd^12+} & MSD & RMSD \\
\hline

DKS/TDA & 11.83 & 13.42 & 15.21 & 17.12 & 19.27 & 21.61 & 24.18 & 26.98 & 30.03 & 33.34 & & \\
amfX2C/TDA & 11.83 & 13.42 & 15.21 & 17.12 & 19.27 & 21.61 & 24.18 & 26.98 & 30.03 & 33.34 & \bf{ 0.00} & \bf{ 0.00} \\
1eX2C/TDA & 12.46 & 14.11 & 15.94 & 17.93 & 20.14 & 22.56 & 25.20 & 28.08 & 31.21 & 34.61 & \bf{ 0.93} & \bf{ 0.30} \\
SO-ZKS/SCIS & 11.69 & 13.26 & 15.01 & 16.86 & 19.00 & 21.31 & 23.83 & 26.59 & 29.58 & 32.84 & \bf{-0.30} & \bf{ 0.10} \\
SR-ZKS/TDA+SO & 11.20 & 12.70 & 14.37 & 16.18 & 18.19 & 20.39 & 22.79 & 25.40 & 28.24 & 31.31 & \bf{-1.22} & \bf{ 0.41} \\
SR-ZKS/SCIS+SO & 11.18 & 12.67 & 14.34 & 16.11 & 18.12 & 20.31 & 22.70 & 25.30 & 28.12 & 31.18 & \bf{-1.30} & \bf{ 0.44} \\
\hline
\hline
$\Delta$M$_{4,5}$ & \ce{Y^3+} & \ce{Zr^4+} & \ce{Nb^5+} & \ce{Mo^6+} & \ce{Tc^7+} & \ce{Ru^8+} & \ce{Rh^9+} & \ce{Pd^10+} & \ce{Ag^11+} & \ce{Cd^12+} & MSD & RMSD \\
\hline

DKS/TDA &  1.90 &  2.12 &  2.34 &  2.56 &  2.77 &  3.00 &  3.25 &  3.55 &  3.85 &  4.18 & & \\
amfX2C/TDA &  1.90 &  2.12 &  2.34 &  2.56 &  2.77 &  3.00 &  3.25 &  3.54 &  3.84 &  4.18 & \bf{-0.00} & \bf{ 0.00} \\
1eX2C/TDA &  2.57 &  2.87 &  3.17 &  3.47 &  3.78 &  4.10 &  4.46 &  4.88 &  5.30 &  5.76 & \bf{ 1.09} & \bf{ 0.36} \\
SO-ZKS/SCIS &  1.80 &  1.98 &  2.19 &  2.42 &  2.74 &  3.00 &  3.31 &  3.63 &  3.89 &  4.20 & \bf{-0.04} & \bf{ 0.03} \\
SR-ZKS/TDA+SO &  1.79 &  2.01 &  2.25 &  2.50 &  2.77 &  3.06 &  3.37 &  3.73 &  4.07 &  4.48 & \bf{ 0.05} & \bf{ 0.05} \\
SR-ZKS/SCIS+SO &  1.77 &  1.99 &  2.22 &  2.47 &  2.73 &  3.02 &  3.33 &  3.69 &  4.02 &  4.41 & \bf{ 0.01} & \bf{ 0.04} \\
\hline
\hline
$\Delta$N$_{4,5}$ & \ce{Lu^3+} & \ce{Hf^4+} & \ce{Ta^5+} & \ce{W^6+} & \ce{Re^7+} & \ce{Os^8+} & \ce{Ir^9+} & \ce{Pt^10+} & \ce{Au^11+} & \ce{Hg^12+} & MSD & RMSD \\
\hline

DKS/TDA &  9.00 &  9.35 &  9.65 &  9.89 & 10.06 & 10.23 & 10.34 & 10.47 & 10.65 & 11.11 & & \\
amfX2C/TDA &  9.02 &  9.35 &  9.65 &  9.89 & 10.06 & 10.23 & 10.34 & 10.47 & 10.64 & 11.11 & \bf{ 0.00} & \bf{ 0.00} \\
1eX2C/TDA &  9.28 & 10.64 & 11.35 & 11.50 & 12.21 & 12.46 & 12.70 & 12.96 & 13.28 & 15.43 & \bf{ 2.11} & \bf{ 0.74} \\
SO-ZKS/SCIS &  8.76 &  9.08 &  9.44 &  9.78 & 10.02 & 10.38 & 10.83 & 11.18 & 11.63 & 12.09 & \bf{ 0.24} & \bf{ 0.17} \\
SR-ZKS/TDA+SO &  8.54 &  8.87 &  9.28 &  9.66 &  9.96 & 10.40 & 10.94 & 11.37 & 11.85 & 12.34 & \bf{ 0.25} & \bf{ 0.22} \\
SR-ZKS/SCIS+SO &  8.54 &  8.87 &  9.27 &  9.65 &  9.95 & 10.37 & 10.97 & 11.38 & 11.85 & 12.33 & \bf{ 0.24} & \bf{ 0.22} \\
\hline
\hline
\end{tabular}}
\caption{SO splittings, mean signed deviation (MSD), and root mean squared deviation (RMSD) in the X-ray edge for a series of bare, d$^0$ transition metal cations, computed with different relativistic Hamiltonians and approximations. The splittings are calculated with respect to the spectral features with dominant p$_{3/2} \leftrightarrow$ d$_{5/2}$ and p$_{1/2} \leftrightarrow$ d$_{3/2}$ character.}
\label{TAB:edge_splittings}
\end{table*}

On average, the errors for the SR-ZKS/TDA+SO and SR-ZKS/SCIS+SO methods are -0.21 and -0.14 eV for $\Delta$L$_{2,3}$, -1.22 and -1.30 for $\Delta$M$_{2,3}$, 0.05 and 0.01 eV for $\Delta$M$_{4,5}$, and 0.25 and 0.24 eV for $\Delta$N$_{4,5}$, respectively. 

Based on these results, the SR-ZKS/SCIS+SO approach seems to provide the best compromise between cost and accuracy. In order to assess its performance in a more realistic situation, we calculated the L$_{2,3}$, M$_{2,3}$, M$_{4,5}$, and N$_{4,5}$-edge spectra of several cyanometallates covering different portions of the periodic table (Figure \ref{FIG:xas_pert}).

\subsection{Cyanometallates}

\begin{figure}[htbp!]
    \includegraphics[scale=1]{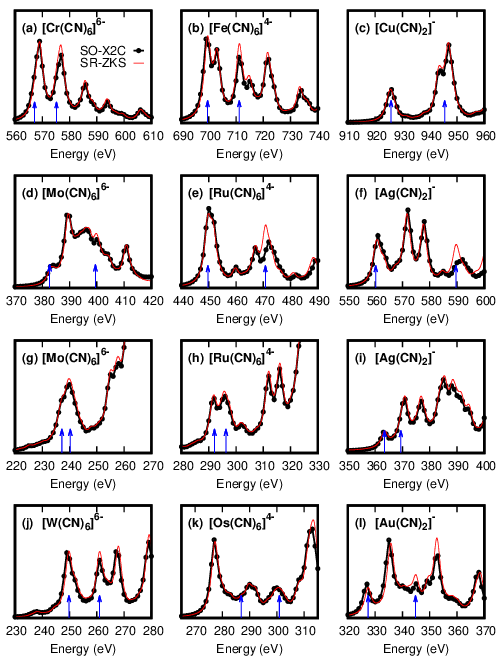}
    \caption{Soft X-ray absorption spectra of cyanometallates at different edges calculated with the SR-ZKS/SCIS+SO (SR-ZKS) and mmfX2C/DR-TDDFT (SO-X2C) methods. A uniform Lorentzian broadening of 1.5 eV was applied to each spectrum. Spectra are shown for the L$_{2,3}$ (a-c), M$_{2,3}$ (d-f), M$_{4,5}$ (g-i), and N$_{4,5}$ (j-l) edges.}
    \label{FIG:xas_pert}
\end{figure}

Figure \ref{FIG:xas_pert} shows the X-ray absorption spectra calculated using the state-interaction-based SR-ZKS/SCIS+SO (SR-ZKS) approach versus those calculated using the damped-response (DR) TD-DFT framework paired with the molecular mean-field X2C Hamiltonian (SO-X2C) of Konecny {\it et al.}\cite{konecny2023exact}  as a reference.   

As one can observe, the SR-ZKS/SCIS+SO spectra reproduces the mmfX2C/DR-TDDFT remarkably well, given the approximations made in the former. Small discrepancies are visible in the M$_{2,3}$-edge spectra of \ce{[Ru(CN)6]^{4-}} (Figure \ref{FIG:xas_pert}e) and \ce{[Ag(CN)2]^{-}} (Figure \ref{FIG:xas_pert}f). These discrepancies are a direct result of the consistent underestimation of the $\Delta$M$_{2,3}$ splittings by the SR-ZKS approach, causing the M$_2$ and M$_3$ features to overlap more than they should. For instance, the feature near 470 eV in the \ce{[Ru(CN)6]^{4-}} is in fact two peaks corresponding to excitations of dominant 3p$_{3/2} \to$ 5d and 3p$_{1/2} \to$ 4d characters. In the SO-X2C calculation, these features have a slightly larger energy separation leading to a broader, lower intensity feature, whereas in SR-ZKS, the features appear as a single peak of higher intensity. A similar observation can be made for near 590 eV for \ce{[Ag(CN)2]^{-}}. 

The peaks chosen to measure the SO splitting in the X-ray edges are marked as blue arrows in Figure \ref{FIG:xas_pert}. Here we chose the lowest-energy features with dominant p$_{3/2} \to$ d and p$_{1/2} \to$ d character as a reference for the X$_{2,3}$-edge spectra, and the lowest-energy features with dominant d$_{5/2} \to$ p and d$_{3/2} \to$ p character as a reference for the X$_{4,5}$-edge spectra, where X correspond to the L, M, or N edges.

An important observation to be made is that the strongest features in the spectra do not always coincide with the metal-centered p$\leftrightarrow$d features. A clear illustration of this can be seen in the case of \ce{[Cr(CN)6]^{6-}}, where the metal centered 2p$_{3/2} \to$ 4d (e$_g$) appears around 567 eV next to a stronger metal-to-ligand 2p$_{3/2} \to$ $\pi_L^*$ (t$_{2g}$) feature at 569 eV. For \ce{[Fe(CN)6]^{4-}}, on the other hand, these features are better separated at 700 and 703 eV, respectively, with the metal-centered absorption having slightly higher intensity. By further inspecting the orbital contributions to each excitation, one can conclude that the larger overlap between these features observed in the Cr L-edge leads to mixing between the two states and consequently to an intensity transfer from the 2p$_{3/2} \to$ 4d (e$_g$) to the 2p$_{3/2} \to$ $\pi_L^*$ (t$_{2g}$) features. 

Furthermore, one can observe that the Mo M$_{2,3}$-edge in \ce{[Mo(CN)6]^{6-}} and Os N$_{4,5}$-edge in \ce{[Os(CN)6]^{4-}} appear at the same energy window as the ligand K-edge, which dominate the spectra. This is clearly obvious in the Os N-edge case, where the simulated spectrum closely resembles the C K-edge spectrum expected for unsaturated organic molecules \cite{nascimento2017simulation}. These observations should be taken into consideration when designing an experiment to probe these edges due to overlapping energies.

\section{Conclusions}
\label{ConclusionSection}

We explored the applicability of a linear-response treatment for a ZKS Hamiltonian, based on the relativistic model potential approach of van Wüllen \cite{van1998molecular,van2005accurate}, to simulate soft X-ray near-edge absorption spectra in closed-shell transition metal systems. Our study demonstrated that spin-orbit coupling effects can be efficiently incorporated using the state-interaction approach, which combines singlet and triplet CIS-like states from scalar-relativistic calculations. We examined two approximations—one including the exchange-correlation contribution in the response equations and one without it. A key advantage of the model potential approach is its compatibility with standard, contracted basis sets, thereby avoiding the high computational costs associated with uncontracted basis sets in fully relativistic calculations.

Our results show that this method accurately reproduces spin-orbit splittings across the periodic table. For bare transition metal cations, the predicted X-ray edge splittings align well with those from 4-component DKS methods. Additionally, for a series of cyanometallates, the state-interaction ZKS approach yields near-edge X-ray absorption spectra that closely agree with results from state-of-the-art X2C methods, all while significantly reducing computational cost.

While not intended to replace more rigorous methods, this quasi-relativistic approach offers a practical alternative for efficiently computing core-level spectra in large, closed-shell transition metal complexes. It is particularly advantageous in cases where full relativistic calculations become computationally prohibitive. Moreover, this method is especially useful for capturing a large number of excited states, making it well-suited for applications such as resonant inelastic X-ray scattering simulations.

\vspace{5pt}
{\bf Acknowledgments}

This work was supported by the National Science Foundation under the CAREER grant No. CHE-2337902 (S.~P, M.~A.~D, and D.~R.~N). N.~G. acknowledges support by the U.S. Department of Energy, Office of Science, Office of Basic Energy Sciences through the Condensed Phase and Interfacial Molecular Science Program of the Division of Chemical Sciences, Geosciences, and Biosciences of the U.S. Department of Energy (DOE) at Pacific Northwest National Laboratory under FWP 80818. This research benefited from computational resources provided by the University of Memphis High-Performance Computing Facility.

\bibliography{main}

\end{document}